\documentclass[journal=jctcce,type=communication]{achemso}
\setkeys{acs}{maxauthors=0,articletitle=true}

\usepackage{achemso}
\usepackage{graphics}
\usepackage{amssymb,amsfonts}
\usepackage{graphicx}
\usepackage[table]{xcolor}
\usepackage{booktabs}
\usepackage{multirow}
\usepackage{caption}
\usepackage{amsmath}
\usepackage{amsopn}
\usepackage{siunitx}
\usepackage{bm}
\usepackage{color}

\author{Janus J. Eriksen}
\email{janus.eriksen@bristol.ac.uk}
\affiliation{School of Chemistry, University of Bristol, Cantock's Close, Bristol BS8 1TS, United Kingdom}
\author{Tyler A. Anderson}
\affiliation{Laboratory of Atomic and Solid State Physics, Cornell University, Ithaca, New York 14853, USA}
\author{J. Emiliano Deustua}
\affiliation{Department of Chemistry, Michigan State University, East Lansing, Michigan 48824, USA}
\author{Khaldoon Ghanem}
\affiliation{Max-Planck-Institut f{\"u}r Festk{\"o}rperforschung, 70569 Stuttgart, Germany}
\author{Diptarka Hait}
\affiliation{Kenneth S. Pitzer Center for Theoretical Chemistry, Department of Chemistry, University of California, Berkeley, California 94720, USA}
\alsoaffiliation{Chemical Sciences Division, Lawrence Berkeley National Laboratory, Berkeley, California 94720, USA}
\author{Mark R. Hoffmann}
\affiliation{Chemistry Department, University of North Dakota, Grand Forks, North Dakota 58202-9024, USA}
\author{Seunghoon Lee}
\affiliation{Division of Chemistry and Chemical Engineering, California Institute of Technology, Pasadena, California 91125, USA}
\author{Daniel S. Levine}
\affiliation{Kenneth S. Pitzer Center for Theoretical Chemistry, Department of Chemistry, University of California, Berkeley, California 94720, USA}
\author{Ilias Magoulas}
\affiliation{Department of Chemistry, Michigan State University, East Lansing, Michigan 48824, USA}
\author{Jun Shen}
\affiliation{Department of Chemistry, Michigan State University, East Lansing, Michigan 48824, USA}
\author{Norm M. Tubman}
\affiliation{Kenneth S. Pitzer Center for Theoretical Chemistry, Department of Chemistry, University of California, Berkeley, California 94720, USA}
\author{K. Birgitta Whaley}
\affiliation{Kenneth S. Pitzer Center for Theoretical Chemistry, Department of Chemistry, University of California, Berkeley, California 94720, USA}
\author{Enhua Xu}
\affiliation{Graduate School of Science, Technology, and Innovation, Kobe University, 1-1 Rokkodai-cho, Nada-ku, Kobe 657-8501, Japan}
\author{Yuan Yao}
\affiliation{Laboratory of Atomic and Solid State Physics, Cornell University, Ithaca, New York 14853, USA}
\author{Ning Zhang}
\affiliation{Beijing National Laboratory for Molecular Sciences, Institute of Theoretical and Computational Chemistry, College of Chemistry and Molecular Engineering, Peking University, Beijing 100871, China}
\author{Ali Alavi}
\email{a.alavi@fkf.mpg.de}
\affiliation{Max-Planck-Institut f{\"u}r Festk{\"o}rperforschung, 70569 Stuttgart, Germany}
\alsoaffiliation{Department of Chemistry, University of Cambridge, Cambridge CB2 1EW, United Kingdom}
\author{Garnet Kin-Lic Chan}
\email{gkc1000@gmail.com}
\affiliation{Division of Chemistry and Chemical Engineering, California Institute of Technology, Pasadena, California 91125, USA}
\author{Martin Head-Gordon}
\email{mhg@cchem.berkeley.edu}
\affiliation{Kenneth S. Pitzer Center for Theoretical Chemistry, Department of Chemistry, University of California, Berkeley, California 94720, USA}
\alsoaffiliation{Chemical Sciences Division, Lawrence Berkeley National Laboratory, Berkeley, California 94720, USA}
\author{Wenjian Liu}
\email{liuwj@sdu.edu.cn}
\affiliation{Qingdao Institute for Theoretical and Computational Sciences, Shandong University, Qingdao, Shandong 266237, China}
\author{Piotr Piecuch}
\email{piecuch@chemistry.msu.edu}
\affiliation{Department of Chemistry, Michigan State University, East Lansing, Michigan 48824, USA}
\alsoaffiliation{Department of Physics and Astronomy, Michigan State University, East Lansing, Michigan 48824, USA}
\author{Sandeep Sharma}
\email{sandeep.sharma@colorado.edu}
\affiliation{Department of Chemistry, The University of Colorado at Boulder, Boulder, Colorado 80302, USA}
\author{Seiichiro L. Ten-no}
\email{tenno@garnet.kobe-u.ac.jp}
\affiliation{Graduate School of Science, Technology, and Innovation, Kobe University, 1-1 Rokkodai-cho, Nada-ku, Kobe 657-8501, Japan}
\author{C. J. Umrigar}
\email{cyrusumrigar@gmail.com}
\affiliation{Laboratory of Atomic and Solid State Physics, Cornell University, Ithaca, New York 14853, USA}
\author{J{\"u}rgen Gauss}
\email{gauss@uni-mainz.de}
\affiliation{Department Chemie, Johannes Gutenberg-Universit{\"a}t Mainz, Duesbergweg 10-14, 55128 Mainz, Germany}

\title[TITLE]{The Ground State Electronic Energy of Benzene}

\begin{document}

\newpage

%
%
\begin{abstract}

We report on the findings of a blind challenge devoted to determining the frozen-core, full configuration interaction (FCI) ground state energy of the benzene molecule in a standard correlation-consistent basis set of double-$\zeta$ quality. As a broad international endeavour, our suite of wave function-based correlation methods collectively represents a diverse view of the high-accuracy repertoire offered by modern electronic structure theory. In our assessment, the evaluated high-level methods are all found to qualitatively agree on a final correlation energy, with most methods yielding an estimate of the FCI value around $-863$ m$E_{\text{H}}$. However, we find the root-mean-square deviation of the energies from the studied methods to be considerable (1.3 m$E_{\text{H}}$), which in light of the acclaimed performance of each of the methods for smaller molecular systems clearly displays the challenges faced in extending reliable, near-exact correlation methods to larger systems. While the discrepancies exposed by our study thus emphasize the fact that the current state-of-the-art approaches leave room for improvement, we still expect the present assessment to provide a valuable community resource for benchmark and calibration purposes going forward.

\end{abstract}

\newpage

%

%
\section*{TOC Graphic}
\begin{figure}[ht]
\begin{center}
\includegraphics[width=\textwidth]{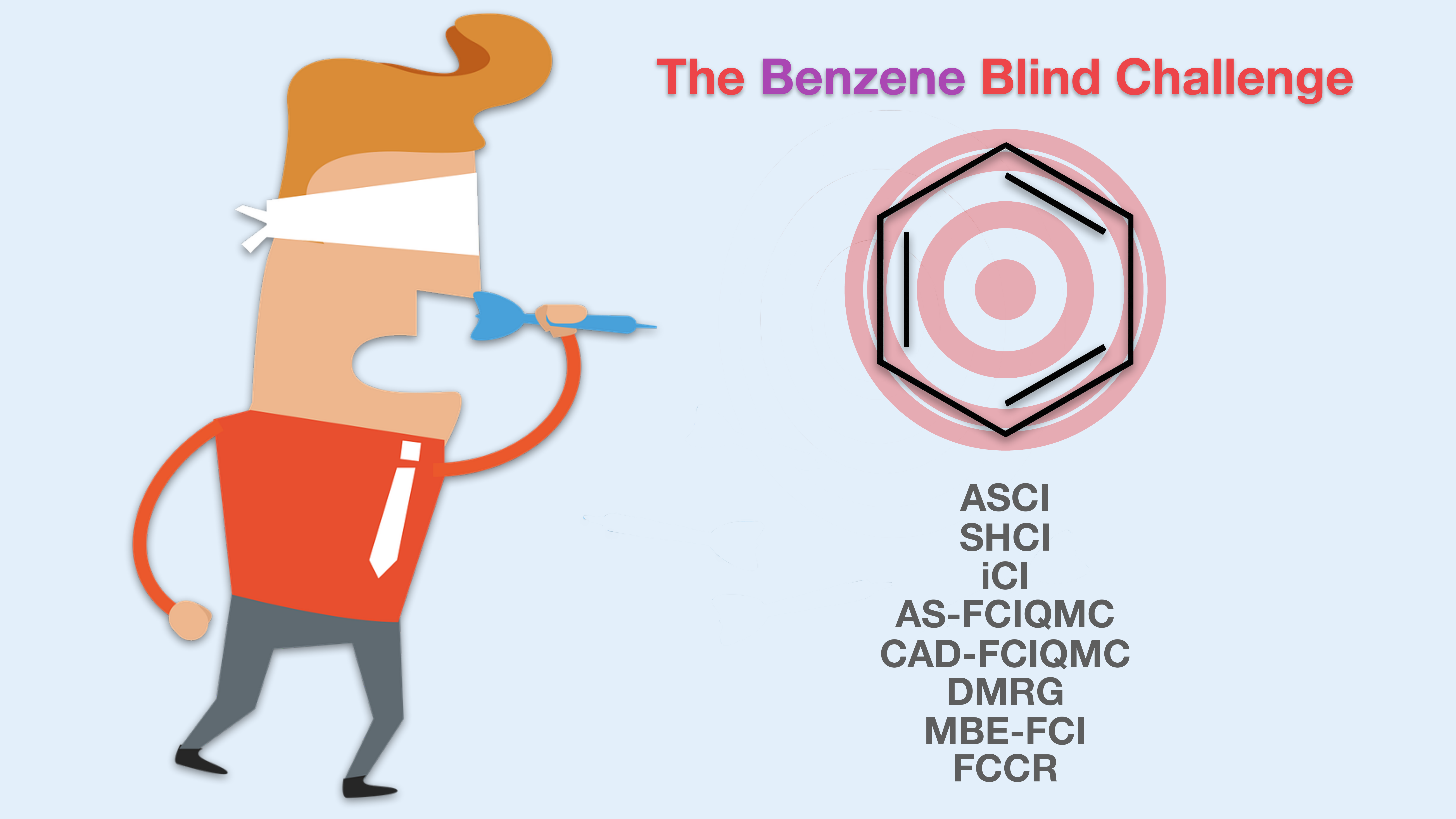}
\label{toc_fig}
\end{center}
\end{figure}
%

%

%
\section*{Keywords}

Benzene; FCI; ASCI; SHCI; iCI; AS-FCIQMC; DMRG; MBE-FCI; FCCR; CAD-FCIQMC

\newpage

%
%

At first glance, the electronic structure of the benzene molecule is deceptively simple. Initially proposed by Kekul{\'e} in the second half of the 19th century~\cite{kekule_benzene_bull_soc_chim_paris_1865,kekule_benzene_liebigs_annalen_1866}, the depiction of benzene as consisting of an alternating pattern of single and double bonds between degenerate carbon atoms was radically novel for its time. Popularly ascribed to a vivid dream of a serpent biting its own tail, the original conjugated structure was soon nuanced in favour of a more balanced, $D_{6\text{h}}$-symmetric resonance picture of benzene~\cite{hueckel_benzene_z_phys_1931,pauling_wheland_benzene_jcp_1933}. However, studies of the finer details of its electronic structure continue to be in vogue to this day~\cite{cooper_gerratt_raimondi_benzene_nature_1986,harcourt_benzene_nature_1987,messmer_schultz_benzene_nature_1987,roos_casscf_benzene_cpl_1992,roos_caspt2_benzene_tca_1995,roos_caspt2_benzene_jcp_2000,gauss_stanton_cc_benzene_jpca_2000,schmidt_benzene_nat_commun_2020}, and an account of its intra- as well as intermolecular physical effects remains a key constraint on a great number of {\textit{ab initio}} simulations in the field of computational (bio-)chemistry~\cite{cacelli_tani_benzene_sim_jacs_2004,cockroft_urch_pi_stacking_jacs_2005,tauer_sherrill_benzene_dimer_jpca_2005,mignon_geerlings_pi_stacking_dna_nuc_acids_res_2005,matta_boyd_pi_stacking_dna_jpcb_2006,schmid_benzene_mof5_sim_angew_chem_2007,geng_sherrill_indole_benzene_sim_jpca_2010,villa_van_der_vegt_benzene_sim_jctc_2010,fu_tian_benzene_sim_jctc_2011,martinez_iverson_pi_stacking_chem_sci_2012}. Even more so, benzene---alongside, for instance, water---may easily be named among the members of an exclusive subset of molecules which are identifiable by wider parts of the public. Constituting the smallest neutral aromatic system composed purely of carbon and hydrogen atoms, benzene rings are omnipresent throughout most of organic chemistry as recurring and easily recognizable structural leitmotifs, to the extent that its widespread use as a symbol of the biological and chemical sciences has become commonplace in society nowadays.\\

That being said, with its total of six carbon atoms, each bonded to a hydrogen atom, benzene has so far been deemed too big to allow for a truly high-level description of its electronic wave function. Even in the modest cc-pVDZ basis set~\cite{dunning_1_orig}, which is the smallest meaningful one-electron basis for use in correlated calculations, and disregarding the six inner core molecular orbitals (MOs), the many-electron Hilbert space of benzene is still on the order of $10^{35}$ Slater determinants, making an exact diagonalization of the Hamiltonian prohibitively expensive. However, given the availability of scalable computational hardware today and, even more importantly, the extensive array of emerging new methods for yielding near-exact electronic ground state energies, we believe that the time is now ripe for an ambitious attempt at solving the electronic Schr{\"o}dinger equation for the ubiquitous benzene molecule.\\

However trivial a problem it might seem, the quest for a numerically near-exact (i.e., sub-m$E_{\text{H}}$ accuracy) treatment of the electron correlation in benzene is complicated by the sheer scale of the combinatorial problem associated with distributing 30 electrons among 108 orbitals. As an illustrative example, upon traversing up through the standard coupled cluster~\cite{cizek_1,cizek_2,paldus_cizek_shavitt} (CC) hierarchy, satisfactory convergence of the correlation energy cannot be concluded even upon accounting for connected quadruple excitations~\cite{ccsdtq_paper_1_jcp_1991,ccsdtq_paper_2_jcp_1992} (CCSDTQ), which is the highest level of sophistication possible today for systems of this size~\cite{matthews_stanton_ccsdtq_jcp_2015,ncc}. In general, assuming a reasonably dominant Hartree-Fock (HF) solution, CCSDTQ is expected to recover almost all of the remaining discrepancies against exact full configuration interaction~\cite{knowles_handy_fci_cpl_1984,knowles_handy_fci_jcp_1989,olsen_fci_cpl_1990} (FCI) present in lower-level (CCSD~\cite{ccsd_paper_1_jcp_1982} and CCSDT~\cite{ccsdt_paper_1_jcp_1987,*ccsdt_paper_1_jcp_1988_erratum,ccsdt_paper_2_cpl_1988}) CC models~\cite{mest,shavitt_bartlett_cc_book}. In the case of benzene, CCSDT lowers the energy by a full $-36.45$ m$E_{\text{H}}$ over CCSD, while the inclusion of connected quadruple excitations adds an additional $-2.47$ m$E_{\text{H}}$, yielding a total correlation energy of $\Delta E_{\text{CCSDTQ}} = -862.37$ m$E_{\text{H}}$. To put these numbers in perspective, and to probe whether or not convergence fails to be met at the CCSDTQ level of theory, the energy increments from connected quadruply and higher excited clusters in the N$_2$ molecule (at the equilibrium geometry) have previously been found to be $-1.61$ m$E_{\text{H}}$ and $-0.23$ m$E_{\text{H}}$, respectively~\cite{chan_bond_break_n2_jcp_2004}. Assuming, for the sake of argument, that higher-level correlation effects are of the same relative order in benzene, the final correlation energy might be estimated at about $\Delta E = -863$ m$E_{\text{H}}$ (by multiplying N$_2$ results by a factor of 3). The main objective in the current work is to move beyond this estimate.\\

In an attempt to substantiate the above projections for what might be expected upon moving toward a higher level of correlation treatment, extended CI wave function expansions have been interpreted for the benzene/cc-pVDZ system by means of a cluster decomposition method~\cite{lehtola_head_gordon_fci_decomp_jcp_2017} (cf. the Supporting Information (SI)), which is analogous to the cluster analysis of the wave function exploited in externally corrected CC approaches~\cite{paldus_eccc_pra_1984,stolarczyk_eccc_cpl_1994,paldus_eccc_tca_1994_1,piecuch_paldus_eccc_pra_1996,paldus_eccc_ijqc_1997,paldus_eccc_jcp_1997,malrieu_paldus_cipsi_eccc_jcp_1999,paldus_eccc_jcp_2006,paldus_eccc_jmc_2017}. On the whole, these results appear to indicate that most of the quadruply (and higher) excited determinants in the FCI wave function stem from disconnected clusters, suggesting that the inclusion of connected quintuples, hextuples, etc., in CC theory is relatively insignificant in comparison, although the above estimate of the FCI correlation energy indicates that higher--than--quadruply excited clusters may play a nontrivial role when trying to obtain results accurate to within fractions of a m$E_{\text{H}}$. The accurate determination of the electronic ground state energy of benzene hence becomes more than an exercise of mere academic interest. Not only does the benzene molecule constitute a challenging test application to push the limits of contemporary, near-exact electronic structure theory, but our results will further allow us to scrutinize the preliminary observations discussed above, namely, to what extent higher-order connected excitations contribute to the FCI correlation energy for an archetypal, medium-sized molecular system with no obvious indications of strong electron correlations.\\

The present study thus aligns itself with the recent series of meticulous benchmark studies from the {\textit{Simons Collaboration on the Many-Electron Problem}} concerned with model systems and small transition-metal species~\cite{simons_collab_hubbard_prx_2015,simons_collab_bond_break_hydrogen_prx_2017,simons_collab_electronic_hamiltonians_prx_2020}. However, as opposed to these earlier assessments, we have conducted the present study as a blind challenge with one of us (J.G.) responsible for compiling all results. This was done in an attempt to conduct an unbiased evaluation of the various methods used in the present work, as listed in Table \ref{abbreviations_table}. Not only are the results of our study bound to prove valuable to future benchmarks and for the calibration of future methods across most of electronic structure theory, but the scatter of the resulting correlation energies further admits a direct assessment of state-of-the-art approaches nearly a century on from the dawn of modern quantum mechanics~\cite{schroding_wave_equation_1926,born_oppenheimer_approx_1927,dirac_quantum_mechanics_book_1930}, in particular in terms of performance transferability in moving from small- to modest-sized molecular compounds. We will herein refrain from passing judgement on what a tolerable error with respect to our FCI target amounts to, since the accuracy of any calculation needs to be weighed against the computational effort required to obtain a particular result to paint a full picture. As such, we will report our findings below in an intentionally neutral tone, leaving most interpretations of the data to the reader.\\

\begin{table}[ht]
\begin{center}
\caption{Abbreviations used for the methods included in the blind challenge.}
\label{abbreviations_table}
\begin{tabular}{l|l|c}
\hline
\multicolumn{1}{c|}{Acronym} & \multicolumn{1}{c|}{Method} & \multicolumn{1}{c}{References} \\
\hline\hline
ASCI & Adaptive Sampling CI & \citenum{tubman_whaley_selected_ci_jcp_2016,tubman_whaley_selected_ci_jctc_2020,tubman_whaley_selected_ci_pt_arxiv_2018,hait_head_gordon_cc_3d_metals_jctc_2019,levine_head_gordon_selected_ci_casscf_jctc_2020} \\
SHCI & Semistochastic Heat-Bath CI & \citenum{petruzielo_umrigar_spmc_prl_2012,holmes_umrigar_heat_bath_fock_space_jctc_2016,holmes_umrigar_heat_bath_ci_jctc_2016,sharma_umrigar_heat_bath_ci_jctc_2017,smith_sharma_heat_bath_casscf_jctc_2017,holmes_sharma_heat_bath_ci_excited_states_jcp_2017,li_sharma_umrigar_heat_bath_ci_jcp_2018} \\
iCI & Iterative CI with Selection & \citenum{liu_hoffmann_sds_tca_2014,liu_hoffmann_ici_jctc_2016,liu_hoffmann_sdspt2_mp_2017,liu_hoffmann_ici_jctc_2020} \\
AS-FCIQMC & Adaptive-Shift FCI Quantum Monte Carlo & \citenum{booth_alavi_fciqmc_jcp_2009,cleland_booth_alavi_fciqmc_jcp_2010,booth_alavi_fciqmc_nature_2013,booth_alavi_fciqmc_jcp_2017,ghanem_alavi_fciqmc_jcp_2019} \\
DMRG & Density Matrix Renormalization Group & \citenum{white_dmrg_prl_1992,white_dmrg_prb_1993,white_martin_dmrg_jcp_1999,mitrushenkov_palmieri_dmrg_jcp_2001,legeza_hess_dmrg_prb_2003,chan_head_gordon_dmrg_jcp_2002,chan_sharma_dmrg_review_arpc_2011,sharma_chan_dmrg_2012,chan_dmrg_review_jcp_2015,wouters_dmrg_review_epjd_2014,yanai_dmrg_review_ijqc_2015,knecht_reiher_dmrg_review_chimica_2016} \\
MBE-FCI & Many-Body Expanded FCI & \citenum{eriksen_mbe_fci_jpcl_2017,eriksen_mbe_fci_weak_corr_jctc_2018,eriksen_mbe_fci_strong_corr_jctc_2019,eriksen_mbe_fci_general_jpcl_2019} \\
FCCR & Full CC Reduction & \citenum{ten_no_fccr_prl_2018} \\
CAD-FCIQMC & Cluster-Analysis-Driven FCIQMC & \citenum{piecuch_monte_carlo_cc_jcp_2018,piecuch_cad-fciqmc-unpublished} \\
\hline
\end{tabular}
\vspace{-0.6cm}
\end{center}
\end{table}
For the sake of brevity, technical details on the evaluated methods listed in Table \ref{abbreviations_table} and the detailed results obtained in our calculations are collected in the SI. Here, we will only briefly compare the methods on the basis of their common traits and differences. The adaptive sampling CI~\cite{tubman_whaley_selected_ci_jcp_2016,tubman_whaley_selected_ci_jctc_2020,tubman_whaley_selected_ci_pt_arxiv_2018,hait_head_gordon_cc_3d_metals_jctc_2019,levine_head_gordon_selected_ci_casscf_jctc_2020} (ASCI), semistochastic heat-bath CI~\cite{petruzielo_umrigar_spmc_prl_2012,holmes_umrigar_heat_bath_fock_space_jctc_2016,holmes_umrigar_heat_bath_ci_jctc_2016,sharma_umrigar_heat_bath_ci_jctc_2017,smith_sharma_heat_bath_casscf_jctc_2017,holmes_sharma_heat_bath_ci_excited_states_jcp_2017,li_sharma_umrigar_heat_bath_ci_jcp_2018} (SHCI), and iterative CI with selection~\cite{liu_hoffmann_sds_tca_2014,liu_hoffmann_ici_jctc_2016,liu_hoffmann_sdspt2_mp_2017,liu_hoffmann_ici_jctc_2020} (iCI) methods all belong to a wider class of selected CI (SCI) methods~\cite{malrieu_cipsi_jcp_1973,harrison_selected_ci_jcp_1991,bagus_selected_ci_jcp_1991,wulfov_selected_ci_cpl_1996,stampfuss_wenzel_selected_ci_jcp_2005,schriber_evangelista_selected_ci_jcp_2016,schriber_evangelista_adaptive_ci_jctc_2017,loos_cipsi_jcp_2018,loos_jacquemin_cipsi_exc_state_jctc_2018,fales_koch_martinez_rrfci_jctc_2018,lu_coord_descent_fci_jctc_2019,berkelbach_fci_fri_jctc_2019,blunt_sci_fciqmc_jcp_2019}, which approximate the full linear expansion of the FCI wave function by selecting only important determinants in conjunction with perturbative corrections to account for any residual correlation. The FCI Quantum Monte Carlo~\cite{booth_alavi_fciqmc_jcp_2009,cleland_booth_alavi_fciqmc_jcp_2010,booth_alavi_fciqmc_nature_2013,booth_alavi_fciqmc_jcp_2017} (FCIQMC) method offers another approach for sampling the wave function, namely, a stochastic QMC propagation of the wave function in the many-electron Hilbert space aimed at projecting out the FCI ground state. The FCIQMC method is most often complemented by an initiator approximation ($i$-FCIQMC), but we will here evaluate its most recent version which uses an adaptive shift~\cite{ghanem_alavi_fciqmc_jcp_2019} (AS-FCIQMC) to mitigate the initiator bias in the wave function sampling. Operating instead using a variational matrix product state Ansatz, density matrix renormalization group~\cite{white_dmrg_prl_1992,white_dmrg_prb_1993,white_martin_dmrg_jcp_1999,mitrushenkov_palmieri_dmrg_jcp_2001,legeza_hess_dmrg_prb_2003,chan_head_gordon_dmrg_jcp_2002,chan_sharma_dmrg_review_arpc_2011,sharma_chan_dmrg_2012,chan_dmrg_review_jcp_2015,wouters_dmrg_review_epjd_2014,yanai_dmrg_review_ijqc_2015,knecht_reiher_dmrg_review_chimica_2016} (DMRG) methods provide an alternative route toward variationally solving the Schr{\"o}dinger equation. DMRG methods reduce the exponential scaling of the above methods with volume to an exponential scaling in the cross-section area. In the recently proposed many-body expanded FCI~\cite{eriksen_mbe_fci_jpcl_2017,eriksen_mbe_fci_weak_corr_jctc_2018,eriksen_mbe_fci_strong_corr_jctc_2019,eriksen_mbe_fci_general_jpcl_2019} (MBE-FCI) method, the FCI correlation energy (without recourse to the electronic wave function) is decomposed and solved for. By enforcing a strict partitioning of the complete set of MOs into a reference and an expansion space, the residual correlation in the latter of these two spaces is recovered by means of an MBE in the spatial MOs of a given system. Finally, two methods founded on CC theory have been evaluated. In the full CC reduction~\cite{ten_no_fccr_prl_2018} (FCCR) method, cluster projection manifolds and commutator expressions for higher-level excitations are systematically reduced in order to optimally exploit the sparsity of the FCI wave function, as recast using the CC Ansatz. Alternatively, one can use the semistochastic cluster-analysis-driven FCIQMC (CAD-FCIQMC) approach~\cite{piecuch_monte_carlo_cc_jcp_2018,piecuch_cad-fciqmc-unpublished}, in which, in the spirit of the externally corrected CC methods~\cite{paldus_eccc_pra_1984,stolarczyk_eccc_cpl_1994,paldus_eccc_tca_1994_1,piecuch_paldus_eccc_pra_1996,paldus_eccc_ijqc_1997,paldus_eccc_jcp_1997,malrieu_paldus_cipsi_eccc_jcp_1999,paldus_eccc_jcp_2006,paldus_eccc_jmc_2017}, the singly and doubly excited clusters are iterated in the presence of their three- and four-body counterparts extracted from FCIQMC (cf. Refs. \citenum{piecuch_monte_carlo_cc_prl_2017,piecuch_monte_carlo_eom_cc_jcp_2019,piecuch_monte_carlo_eom_cc_mp_2020} for other ways of merging stochastic FCIQMC or CC Monte Carlo~\cite{thom_stochast_cc_prl_2010,franklin_thom_stochast_cc_jcp_2016} with the deterministic CC framework).\\

Among the evaluated methods, a few make use of extrapolations. In the methods that involve a perturbative correction as an integral component on top of a variational calculation (ASCI, SHCI, and iCI), final results may be extrapolated by systematically reducing the portion of the total correlation energy accounted for by second-order perturbation theory. In the case of DMRG, extrapolations may be performed towards an infinite bond dimension estimate. In order to isolate the effect of extrapolation from the bare methods themselves, we will present both the unextrapolated and extrapolated results. On the other hand, MBE-FCI and AS-/CAD-FCIQMC make no use of extrapolations of any kind. The FCCR method may also be augmented by either the Epstein-Nesbet~\cite{epstein_phys_rev_1926,nesbet_proc_1955} or M{\o}ller-Plesset~\cite{mp2_phys_rev_1934} formulations of perturbation theory, and while no extrapolations may be directly drawn from individual FCCR calculations (except for the most recent variant of the theory, cf. the SI), a final result may be derived using the average of these perturbative corrections in combination with adjustments for the internal thresholds.\\

Besides the methods listed in Table \ref{abbreviations_table}, one additional, complementary result has previously been reported in the literature using the same molecular geometry~\cite{sauer_thiel_cc3_benchmark_jcp_2008}, namely, $i$-FCIQMC~\cite{blunt_fciqmc_jctc_2019}, augmented by perturbation theory~\cite{blunt_fciqmc_jcp_2018}. It should be mentioned that none of the methods examined in our study are variational, as those that are formulated on top of selected CI and DMRG theory have lost this feature upon being corrected by perturbation theory or extrapolated towards an infinite bond dimension, respectively. In the case of AS- and $i$-FCIQMC, one loses a variational bound through stochastic wave function samplings followed by blocking analyses and the use of the projected form of the correlation energy expression rather than an expectation value. FCCR and CAD-FCIQMC do not have a bound as they are based on CC theory, and MBE-FCI is nonvariational due to its expansion in terms of increments.\\

\begin{figure}[ht]
\begin{center}
\includegraphics[width=\textwidth]{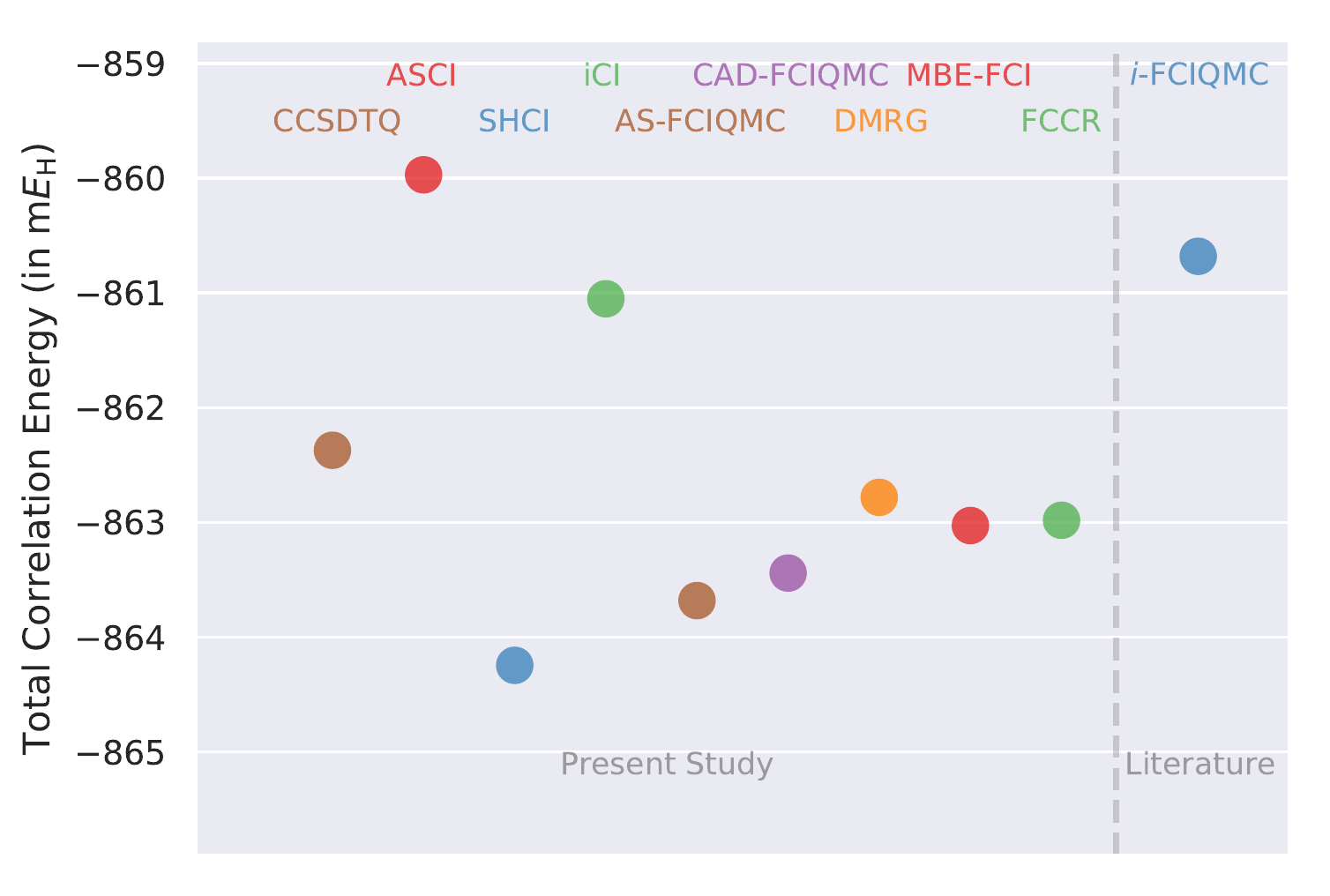}
\caption{Frozen-core C$_6$H$_6$/cc-pVDZ correlation energies for the methods of Table \ref{abbreviations_table} alongside CCSDTQ and $i$-FCIQMC~\cite{blunt_fciqmc_jctc_2019}. For the additional results obtained after the blind test was completed, see the SI.}
\label{results_fig}
\end{center}
\vspace{-0.6cm}
\end{figure}
The main results of our study are summarized in Figure \ref{results_fig} (with the underlying numerical data tabulated in the SI). No error bars are provided given that these are derived differently in the various methods. While our pool of results is too limited to allow for in-depth statistics to be computed from it (besides a mean value, $\mu$, and a standard deviation, $\sigma$), a number of observations may still be made. In the following, we will make use of m$E_{\text{H}}$ as the unit for reporting correlation energies in order to accentuate differences (recalling that 1 m$E_{\text{H}}$ corresponds to $2.6$ kJ/mol).\\

Our key observations can be summarized as follows: {\bf{(i)}} The majority of the methods evaluated in the present work yield a larger correlation energy (in absolute value) than that of the CCSDTQ method, in agreement with the general notion that high-level CC methods, although not bounded by the variational theorem~\cite{nesbet_variational_method}, often are so in practice. {\bf{(ii)}} Across the various results, all but those of the three flavours of SCI fall into an interval ranging from $-863.7$ m$E_{\text{H}}$ to $-862.8$ m$E_{\text{H}}$. {\bf{(iii)}} Taking into account the finer details of the ASCI, iCI, and SHCI calculations (cf. the SI), we expect the result of the latter to be more accurate than the former two, as evidenced by the smallest extrapolation distance among these three methods, cf. Table \ref{extrap_dist_table}; these distances ($\Delta E_{\text{dist}}$) are here meant to serve as an indication of the extent to which the individual methods rely on extrapolation procedures. {\bf{(iv)}} The examples of stochastic CI calculations included in Figure \ref{results_fig} ($i$- and AS-FCIQMC) are also observed to disagree with one another, however only by half of that of their deterministic counterparts. AS-FCIQMC, which corrects for the undersampling bias of noninitiator determinants, is expected to be the more accurate of these two. {\bf{(v)}} The extrapolated DMRG result is in good agreement with the remaining methods listed in point {\bf{(ii)}}. In addition, it is observed from Table \ref{extrap_dist_table} to be far less reliant on an extrapolation of the energy than the tested SCI methods. {\bf{(vi)}} Likewise, the CAD-FCIQMC and MBE-FCI results, both of which have not been extrapolated, agree with each other to within $0.4$ m$E_{\text{H}}$. {\bf{(vii)}} Viewing CAD-FCIQMC as a correction to the underlying AS-FCIQMC wave function, calculating the correlation energy by means of the CC rather than the CI Ansatz is observed to slightly reduce the absolute values of the AS-FCIQMC energies, by $0.3$ m$E_{\text{H}}$ for the most accurate AS-FCIQMC instantaneous and averaged wave functions equilibrated using a population of 2 billion walkers. The deterministic CAD-FCIQMC iterations reduce the change in the AS-FCIQMC correlation energies, when increasing the walker population from 1 to 2 billion, by a factor of about 2 (from $1.1$ to $0.5$ m$E_{\text{H}}$, cf. the SI). For AS-FCIQMC, the change in energy is a reflection of the initiator bias (or approximation) in addition to the smaller stochastic error. {\bf{(viii)}} As further discussed in the SI, the FCCR results exhibit a pronounced dependence on the choice of perturbative treatment, giving rise to an intrinsic variance of $5.3$ m$E_{\text{H}}$. However, the final, perturbatively corrected FCCR correlation energy is estimated to lie in close proximity to the remaining non-SCI results. {\bf{(ix)}} To that end, the results of the only four methods which have not been aided by second-order perturbation theory (DMRG, MBE-FCI, as well as AS- and CAD-FCIQMC), are observed to coincide to a reasonable extent, spanning an interval of only $0.9$ m$E_{\text{H}}$.\\

\begin{table}[ht]
\begin{center}
\caption{Extrapolation distances, $\Delta E_{\text{dist}}$ (in m$E_{\text{H}}$), involved in computing the final ASCI, iCI, SHCI, and DMRG results in Fig. \ref{results_fig}. These are defined by the difference between the final computed energy, $\Delta E_{\text{final}}$, and the extrapolated energy, $\Delta E_{\text{extrap.}}$ (final variational energies, that is, in the absence of perturbation theory, are presented as $\Delta E_{\text{var.}}$). For the SCI methods, extrapolations are performed toward the limit of vanishing perturbative correction, while the variational DMRG energy is extrapolated toward an infinite bond dimension. See the SI for results obtained after the blind test was completed.}
\label{extrap_dist_table}
\begin{tabular}{l|r|r|r|r}
\toprule
\multicolumn{1}{c|}{Method} & \multicolumn{1}{c|}{$\Delta E_{\text{var.}}$} & \multicolumn{1}{c|}{$\Delta E_{\text{final}}$} & \multicolumn{1}{c|}{$\Delta E_{\text{extrap.}}$} & \multicolumn{1}{c}{$\Delta E_{\text{dist}}$} \\
\midrule\midrule
ASCI & $-737.1$ & $-835.4$ & $-860.0$ & $-24.6$ \\
iCI & $-730.0$ & $-833.7$ & $-861.1$ & $-27.4$ \\
SHCI & $-827.2$ & $-852.8$ & $-864.2$ & $-11.4$ \\
DMRG & $-859.2$ & $-859.2$ & $-862.8$ & $-3.6$ \\
\midrule
\end{tabular}
\vspace{-0.6cm}
\end{center}
\end{table}
All of the methods evaluated herein are the products of years of intense development, and most of the computed results in Figure \ref{results_fig} have required a considerable amount of computational resources to obtain. Due to its high polynomial scaling and memory requirements, the CCSDTQ model is unlikely to enable near-exact results for molecular systems significantly larger than benzene. Be that as it may, our CCSDTQ result was still obtained using only 5.5k core hours using a single thread on a multicore node equipped with 120 GB of physical memory, indicating that high-level CC theory represents an affordable, yet robust alternative to many of the other methods tested in our study for problems of a similar size and with similar nature of the involved electron correlation. In comparison, the FCCR result in Figure \ref{results_fig} required a total of 0.1M core hours (using 640 parallel processes) across all of the involved calculations, while the extrapolated DMRG result required 0.08M core hours in total, distributed across $100-200$ cores. The DMRG method generally requires a non-negligible amount of memory, on par or greater than the CC requirements above, while these may be reduced somewhat in the FCCR method. The extrapolated ASCI, SHCI, and iCI results were all obtained in parallel, consuming 0.3k, 2.8k, and 1.5k core hours in the process, respectively, thus all offering relatively inexpensive compromises in comparison with some of the other methods in Figure \ref{results_fig}. Again, the memory requirements involved in running the largest possible CI spaces will ultimately hinder their application to significantly larger problem sizes and basis sets. Both the AS-FCIQMC and MBE-FCI results were obtained in a highly parallel manner, but with minimal memory demands in the case of the latter method. In the case of AS-FCIQMC, a total of 0.06M core hours were consumed, distributed over a group of either 100 or 200 multicore nodes, while the MBE-FCI calculation was parallelized over 128 nodes for a total of 1.7M core hours, by far the most expensive of all the evaluated methods. Finally, the CAD-FCIQMC correlation energy was computed in just a few hours on a single node, initialized from the converged AS-FCIQMC solution.\\

In summary, while all of the methods of our assessment yield results in general agreement with one another, the overall low resolution, as exemplified by a substantial standard deviation across our sample set (in excess of $1.3$ m$E_{\text{H}}$), ultimately hinders a precise determination of the FCI correlation energy to within a small fraction of a m$E_{\text{H}}$. That being said, this uncertainty is most likely too pessimistic, and our findings do indeed seem to indicate, taking into account also the post blind-test energies of some of the methods, that the most plausible frozen-core correlation energy---for the current geometry in the cc-pVDZ basis set---is around $-863$ m$E_{\text{H}}$, in accordance with our preliminary estimate in the introduction and earlier projections~\cite{chan_dmrg_science_2014}. On this basis, we are led to conclude that the electronic structure of benzene in its equilibrium geometry is predominantly dynamic in character.\\

More generally, in particular in view of its format as a blind challenge, our findings collectively represent an unbiased assessment of a diverse set of current state-of-the-art methods. As a consequence of the fact that the sophistication and application range of near-exact electronic structure continue to be improved, we end by strongly encouraging the continued benchmarking of future correlation methods aimed at FCI against the results presented here. To that effect, we note that updated ASCI, SHCI, iCI, and FCCR results---made possible solely by improvements to the efficiencies of their implementations or the use of optimized MOs in combination with larger correlation spaces---were submitted following the compilation of the results in Fig. \ref{results_fig}. These results are discussed in the SI. In addition, two sets of results obtained using alternative methods---phaseless auxiliary-field quantum Monte Carlo~\cite{reichman_afqmc_benzene_arxiv_2020} (ph-AFQMC) and CI using a perturbative selection made iteratively~\cite{loos_cipsi_benzene_arxiv_2020} (CIPSI)---have subsequently appeared in the literature as complementary notes to the present work. 

%
%
\section*{Acknowledgments}

The authors thank Dr. Devin A. Matthews of the Southern Methodist University for help with obtaining the CCSDTQ result first reported in Ref. \citenum{eriksen_mbe_fci_general_jpcl_2019}. J.J.E. is grateful to the Alexander von Humboldt Foundation and the Independent Research Fund Denmark for financial support. J.J.E. and J.G. gratefully acknowledge access awarded to the Galileo supercomputer at CINECA (Italy) through the $18^{\text{th}}$ PRACE Project Access Call and the Johannes Gutenberg-Universit{\"a}t Mainz for computing time granted on the MogonII supercomputer. D.H. and M.H.G. were supported by the Director, Office of Science, Office of Basic Energy Sciences, of the U.S. Department of Energy under Contract No. DE-AC02-05CH11231. P.P. and members of his group, J.E.D., I.M., and J.S., acknowledge support by the Chemical Sciences, Geosciences and Biosciences Division, Office of Basic Energy Sciences, Office of Science, U.S. Department of Energy (grant no. DE-FG02-01ER15228 to P.P.). N.Z. and W.L. acknowledge support from the National Natural Science Foundation of China (grant no. 21033001 and 21973054). S.L. and G.K.C. were supported by the U.S. National Science Foundation, via grant no. 1665333. E.X. and S.L.T. thank the financial support from the Japan Society for the Promotion of Science, Grant-in-Aids for Scientific Research (A) (Grant No. JP18H03900). M.R.H. acknowledges the North Dakota University System. S.S. was supported by the U.S. National Science Foundation grant CHE-1800584 and by the Sloan research fellowship. T.A.A. and C.J.U. were supported in part by the U.S. Air Force Office of Scientific Research under grant FA9550-18-1-0095. J.E.D. and Y.Y. acknowledge support from the Molecular Sciences Software Institute, funded by U.S. National Science Foundation grant ACI-1547580. Some of the SHCI computations were performed at the Bridges cluster at the Pittsburgh Supercomputing Center supported by U.S. National Science Foundation grant ACI-1445606. A.A. and K.G. thank the {\texttt{NECI}} developer team and the Max Planck Computing and Data Facility for their continuing work on the {\texttt{NECI}} code.

%
%
\section*{Supporting Information}

Details on all methods and their results are collected in the Supporting Information. Section 1 lists the geometry and Section 2 summarizes the main results. Section 3 provides details on the MBE-FCI method and its results, and similar details are provided for DMRG, AS-FCIQMC, CAD-FCIQMC, SHCI, ASCI, iCI, FCCR, and CCSDTQ in Sections 4 through 11, respectively. Finally, Section 12 presents cluster decompositions of a few SCI wave functions.

\newpage

\providecommand{\latin}[1]{#1}
\makeatletter
\providecommand{\doi}
  {\begingroup\let\do\@makeother\dospecials
  \catcode`\{=1 \catcode`\}=2 \doi@aux}
\providecommand{\doi@aux}[1]{\endgroup\texttt{#1}}
\makeatother
\providecommand*\mcitethebibliography{\thebibliography}
\csname @ifundefined\endcsname{endmcitethebibliography}
  {\let\endmcitethebibliography\endthebibliography}{}

\end{document}